\documentclass[12pt]{article}
\usepackage[dvips]{graphicx}
\usepackage{pdproc}
\usepackage{epsf,axodraw} 

  \makeatletter 
  \def\@cite#1{[#1]} 
  \makeatother    
\begin{document}

\renewcommand{\thefootnote}{\alph{footnote}}
\newcommand{\bea}{\begin{eqnarray}}
\newcommand{\eea}{\end{eqnarray}}
\newcommand{\lsim}
{{\;\raise0.3ex\hbox{$<$\kern-0.75em\raise-1.1ex\hbox{$\sim$}}\;}}
\newcommand{\gsim}
{{\;\raise0.3ex\hbox{$>$\kern-0.75em\raise-1.1ex\hbox{$\sim$}}\;}}

\title{
  Invisible Higgs in theories of large extra dimensions
}

\author{ANINDYA DATTA, KATRI HUITU, \underline{JARI LAAMANEN}$^1$,
and BISWARUP MUKHOPADHYAYA}

\address{ 
$^1$Helsinki Institute of Physics \\
P.O.Box 64, FIN-00014 University of  Helsinki, Finland \\
}

\abstract{
We discuss the possibility of detecting a Higgs boson in electron-positron
collider experiments if large extra dimensions are realized in nature.
In such a case, the Higgs boson can decay invisibly by oscillating
into a graviscalar Kaluza-Klein (KK) tower. We show that the search
for such a Higgs at an $e^+ e^-$ linear collider entails more
complications than are usually thought of in relation to an invisibly
decaying Higgs. 
}

\normalsize\baselineskip=15pt


\section{Introduction} 
\label{sect:intro}
Confirming the Higgs mechanism as the underlying principle of
electroweak symmetry breaking is one of the main goals of upcoming
accelerators.  
The strategy for Higgs search depends on the decay branching fractions
of Higgs ($H$) to different channels.
One possibility is that a Higgs can have a substantial branching ratio
for decay into invisible final states. This possibility has been
underlined in a number of well-motivated theoretical options
(see \cite{Datta:2004jg} and references 1-8 there).
This talk is based on the Ref.~\cite{Datta:2004jg}.

Theories with large extra spatial dimensions have been popular in
recent times.
Broadly two such types of models have been studied so far, namely, the
Arkani-Hamed-Dimopoulos-Dvali (ADD) \cite{ADD} and Randall-Sundrum
(RS) \cite{RS} types. Both of them accomodate extra compact spacelike
dimensions, with gravity propagating in the `bulk', while all the
Standard Model (SM) fields are confined to (3+1) dimensional slices or
`branes'. In ADD-type models, one has a factorizable geometry, where
the projection on the brane leads to a continuum of scalar and tensor
graviton states.
The discussion in this talk is related to the ADD-type of models.

An important feature of these models is that the Higgs boson can mix
with the graviscalars (the 
projection of the graviton on the visible brane).
One can add to the Lagrangian a mixing term
\bea
S=-\xi\int d^4 x\,\sqrt{-g_{ind}}R(g_{ind})H^{\dagger}H,
\eea
where $H$ is the Higgs doublet, $\xi$ is a dimensionless mixing
parameter, $g_{ind}$, the induced metric on the brane and $R$ is the
Ricci scalar.  
One can parameterize such mixing by the following term in the
Lagrangian:
\begin{equation}
{\cal L}_{mix} = \frac{1}{M_P} m_{mix}^3h\sum_n S_n,  
\label{mix}
\end{equation}
where $m_{mix}^3 = 2 \kappa \xi v m_h^2$, $M_P$ is the reduced Planck
mass, and $v$ is the Higgs vacuum expectation value. $\kappa$ can be
expressed in terms of the number of extra dimensions: $\kappa^2 \equiv
{3 (\delta -1)/\delta +2},$
$\delta$ being the number of extra compact spacelike dimensions.

The effects of extra dimensions on the Higgs decay modes are different
in the case of large or small extra dimensions.  
In the case of the large extra dimensions the major effects are due to
the closely spaced KK-levels.  These lead to the possibly effective
`invisible decays' of the Higgs boson, via oscillation into one or the
other state belonging to the quasi-continuous tower of graviscalars
\cite{GRW}. In case of the small extra dimensions the effect comes
from the mixing of the Higgs boson with the single radion \cite{RS}.

\section{Higgs invisible decays with scalar-graviscalar mixing}

One proceeds by considering the Higgs propagator in the flavor basis
and incorporating all the insertions induced by the mixing term.
As has been shown in \cite{GRW}, the effect of having a large
number of real intermediate states inserted leads to the development
of an imaginary term in the propagator. This imaginary part can be
interpreted as an effective decay width entering into the propagator
following the Breit-Wigner scheme \cite{GRW}\footnote{Our calculation
of the width agrees with \cite{BDGW} where the width is twice the
value in \cite{GRW}. See also talk by J.~Gunion in this proceedings.}:
\bea
\Gamma_G=2\pi\kappa^2\xi^2v^2\frac{m_h^{1+\delta}}{M_D^{2+\delta}}
\frac{2\pi^{\delta/2}}{\Gamma(\delta/2)},
\label{osc}
\eea
where $M_D$ is the $(4+\delta)$-dimensional Planck scale (also 
called the string scale).
A Higgs boson has a finite probability (proportional to $\xi^2$) of
oscillating into a the invisible states corresponding to the
graviscalar tower.
The transition is favored when masses of the Higgs and the
corresponding graviscalar are close to each other.

\begin{figure}[t]
\begin{center}
\hspace{-1cm}
\centerline{\hspace*{3em}
\epsfxsize=8cm\epsfysize=8cm
                     \epsfbox{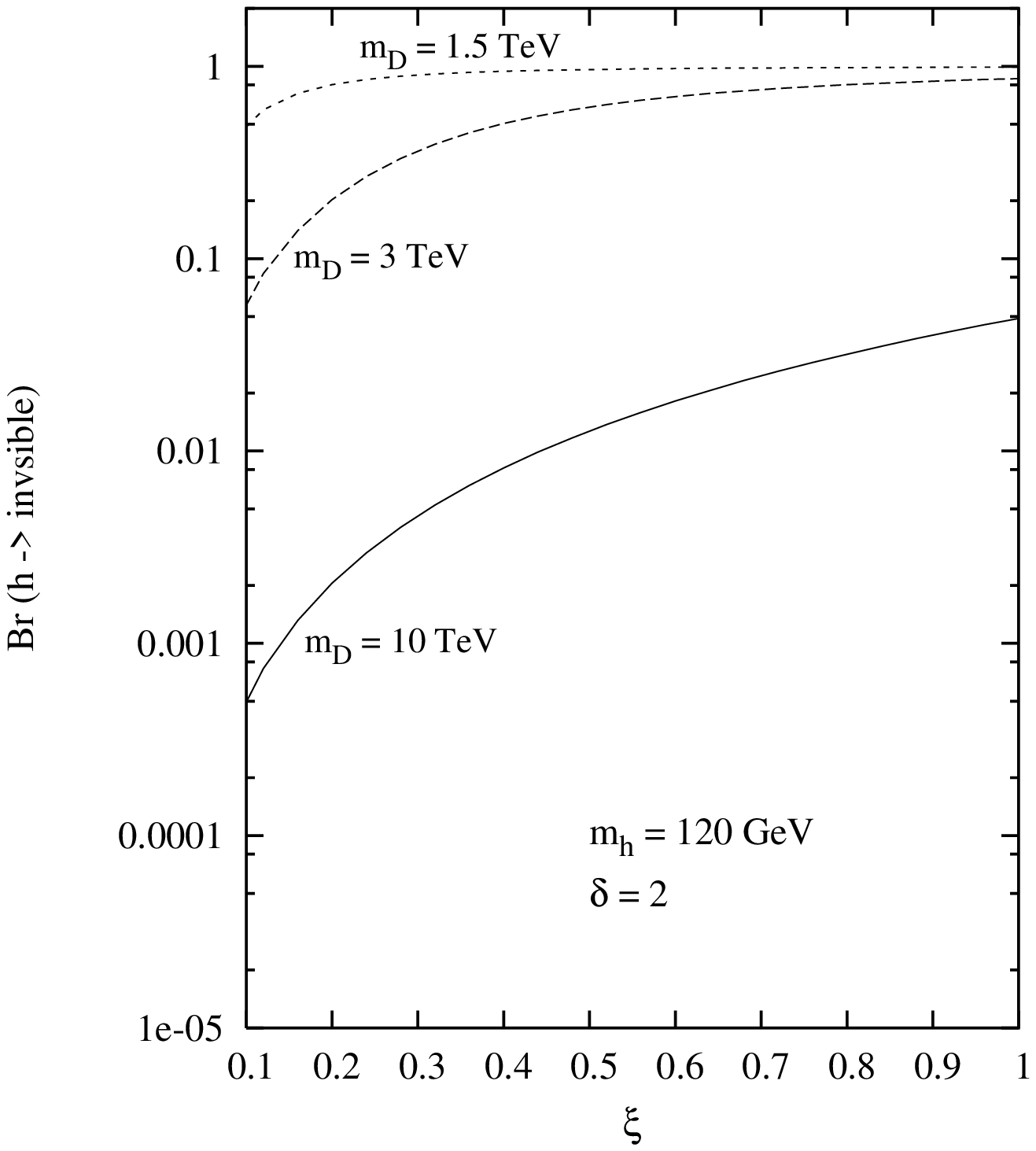}
\epsfxsize=8cm\epsfysize=8cm
                     \epsfbox{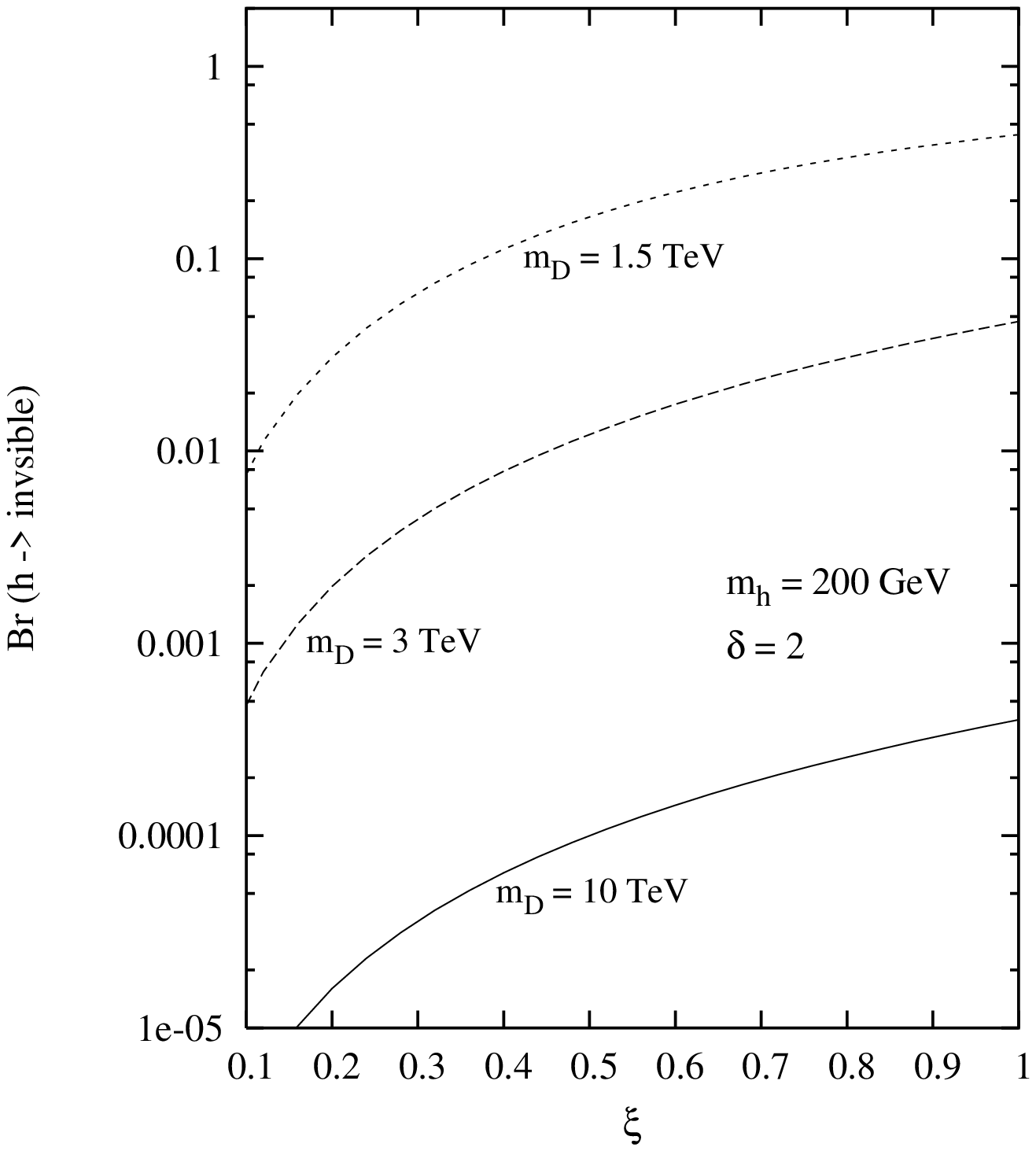}
}
\end{center}

\caption{\label{invBR}The invisible decay branching ratio of the Higgs boson 
as a function of the mixing parameter for two representative values of
Higgs mass and for one number of extra dimensions, $\delta = 2$. }

\end{figure}

Assuming that the graviton KK tower is the only source of the
invisible width in the model, we have plotted in Fig.~\ref{invBR} the
invisible branching ratio as a function of the mixing parameter. Two
masses for the Higgs boson, namely, $m_h=120$ GeV, and $m_h=200$ GeV,
have been used. The plots have been made for $M_D=1.5$, 3, and 10 TeV
and for $\delta=2$.

The effective invisible decay width grows as $m_h ^3$ for $\delta =
2$.  This implies that even for $m_h < 2 m_W$ total Higgs decay width
can be considerably larger than the Standard Model width.
As a consequence, even for a light Higgs boson, Higgs
resonance may not be very sharp.

\section{Detection in $e^+e^-$ collider}

It is evident from the Fig.~\ref{invBR} that the Higgs can have a very
large invisible branching ratio in the scenario considered here.
We investigate the situation at an $e^+e^-$ machine, where the
identification of the recoil mass peak against the $Z$-boson in the
Bjorken process is widely known to be a reliable method of detecting
the Higgs boson\footnote{Possibilities to detect an invisibly decaying
Higgs at next generation $e^+ e^-$ collider have also been discussed
in \cite{ms}, but only in the context of 4-dimensional models where
continuum graviton production does not arise at all.}.  We therefore
concentrate on this process ($ e^+e^-\rightarrow Z(\rightarrow
\mu^+\mu^- )h(\rightarrow {\rm inv})$), also known as the
Higgs-strahlung process, at a linear collider with center of mass
energy of 1 TeV.
The final state we are interested in comprises of a $\mu ^+ \mu ^-$
pair with missing energy/momentum.
We do not use the narrow width approximation, but rather
treat the Higgs boson as a propagator (inclusive of an invisible
width) while calculating the cross-section.  We have also taken into
account the direct graviscalar production alongside.

A similar final state can arise 
also in the production of a
$Z$-boson with towers of graviton (spin-2/spin-0).
In addition to the SM processes $e^+ e^- \rightarrow ZZ/WW
\rightarrow \mu^+ \mu^- + \not{\!\!E}$, we have included
gravitensor production together with a $Z$-boson, both leading to
identical final states.

The $ZZ$ and $Z\nu\bar{\nu}$ backgrounds are eliminated with specific
cuts \cite{Datta:2004jg}.  We find that these criteria work well so
long as the invisible Higgs mass is well below 250 GeV or thereabout.

We also need to worry about filtering out the Higgs effects from
continuum gravitensor contributions.  The first step for this is to
reconstruct the recoiled invariant mass which peaks at the Higgs mass
modulo the Higgs width and detector resolution.
The other important factor is the height of the peak against the
continuum background. It is determined by the Higgs-graviscalar mixing
$\xi$, the same quantity which also determines the Higgs width, making
the width large for $\xi~=~O(1)$.
This causes the invisible decay recoil mass distribution to lose its
sharp character even for a Higgs mass of the order of 120 GeV.

\begin{figure}[t]
\begin{center}
\hspace{-1cm}
\centerline{\hspace*{3em}
\epsfxsize=8cm\epsfysize=8cm
                     \epsfbox{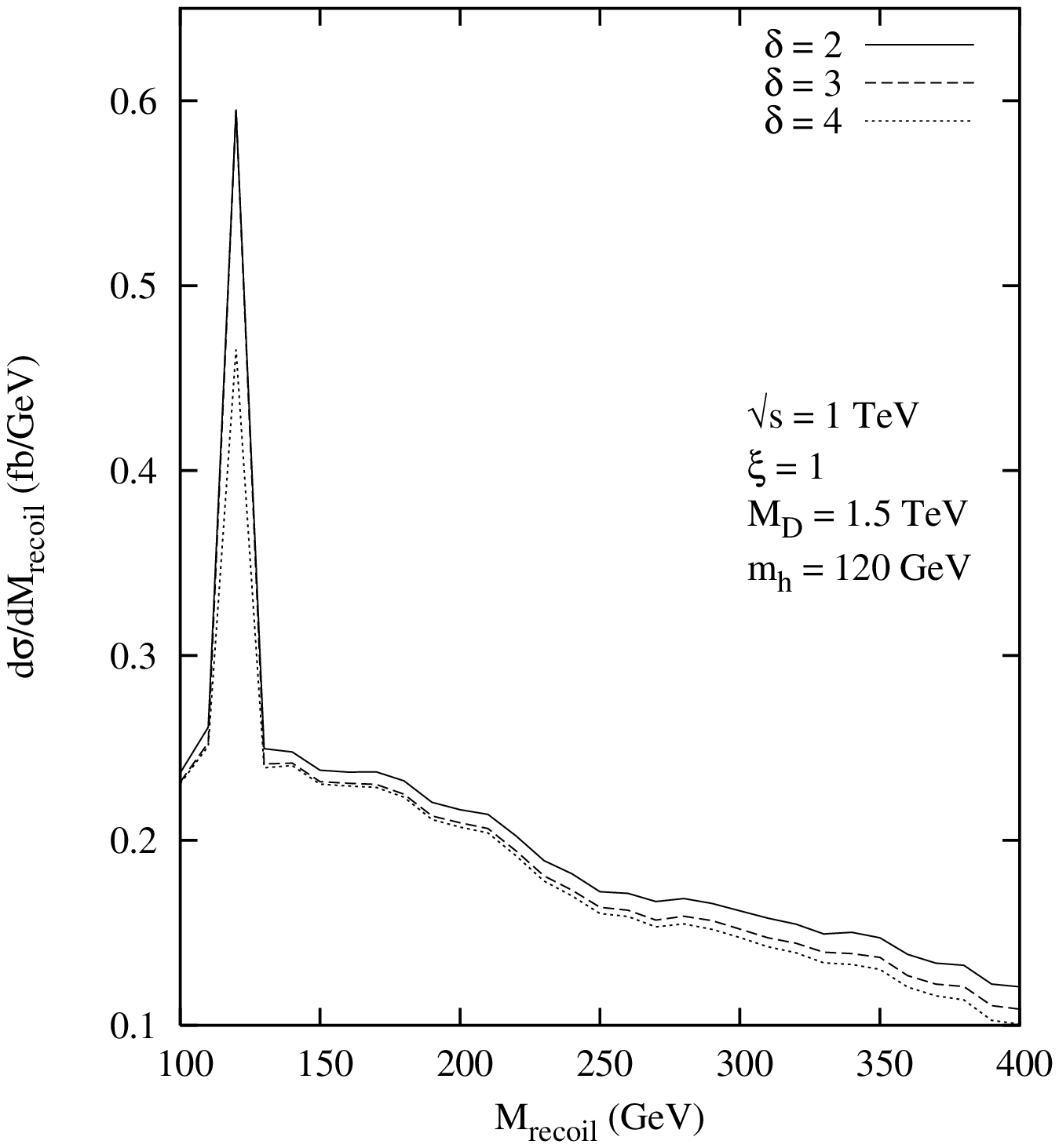}
\epsfxsize=8cm\epsfysize=8cm
                     \epsfbox{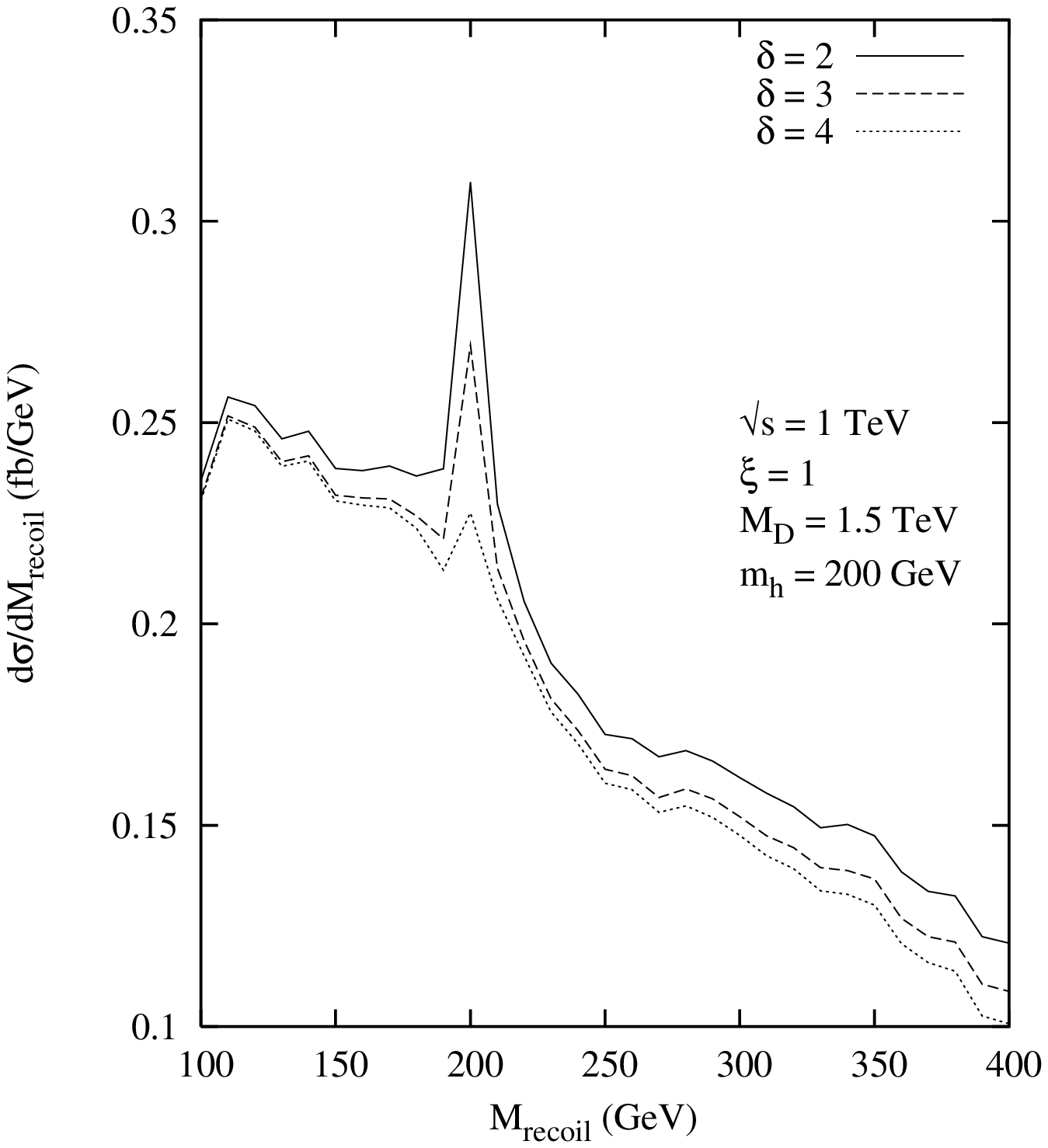}
}
\end{center}

\caption{\label{eecs}Recoil mass distribution for invisible Higgs 
for different values of $\delta$, superposed over the SM and 
gravitensor contributions, for $m_D$ = 1.5 TeV and  $\xi$ = 1.}
\end{figure}

In Fig.~\ref{eecs} we have plotted the recoil mass distribution for
the $\mu^+ \mu^- +\not{\!\!E}$ events including the SM,
gravitensor and invisibly decaying Higgs production.  The results are
presented for three different values of the number of extra
dimensions.  In the two sets of plots, corresponding to Higgs masses
120 GeV (left) and 200 GeV (right), the different degrees of
visibility of the peak is quite obvious.
We have chosen $m_D = 1.5$ TeV, $\xi = 1$ and $\sqrt{s} = 1$ TeV. To
take into account a realistic detector resolution we have assumed a
Gaussian spreading of muon energy
\cite{peskin}. The visibility of the peak for the lower Higgs mass is
largely due to the angular cut \cite{Datta:2004jg}. For a heavier
Higgs, however, reconstruction of the peak appears to be difficult.

\section{Summary}
We have investigated the detection prospects of a Higgs boson that has
mixing with a
tower of graviscalars in a scenario with large compact extra
dimensions. This causes the Higgs to develop an invisible decay
width. In the context of a linear $e^+ e^-$ collider such invisibility
brings in additional problems in reconstructing the Higgs boson as a
recoil mass peak against an identified Higgs boson.



\bibliographystyle{plain}

\end{document}